\documentclass[pra, showpacs, twocolumn]{revtex4}
\usepackage{amssymb}
\usepackage{bm}

\begin{document}
\title
{Semiclassical quantization of non-Hamiltonian dynamical systems
without memory}

\author{E.\ D.\ Vol}

\affiliation{%
B.Verkin Institute for Low Temperature Physics and Engineering
National Academy of Sciences of Ukraine, Lenin av. 47 Kharkov
61103, Ukraine}

\begin{abstract}
We propose a new method of quantization of a wide class of
dynamical systems that originates directly from the equations of
motion. The method is based on the correspondence between the
classical and the quantum Poisson brackets, postulated by Dirac.
This correspondence applied to open (non-Hamiltonian) systems
allows one to point out the way of transition from the quantum
description based on the Lindblad equation to the dynamical
description of their classical analogs by the equations of motion
and vice versa. As the examples of using of the method we describe
the procedure of the quantization of three widely considered
dynamical systems: 1) the harmonic oscillator with friction, 2)
the oscillator with a nonlinear damping that simulates the process
of the emergence of the limit cycle, and 3) the system of two
periodic rotators with a weak interaction that synchronizes their
oscillations. We discuss a possible application of the method for
a description of quantum fluctuations in Josephson junctions with
a strong damping and for the quantization of open magnetic systems
with a dissipation and a pumping.
\end{abstract}

 \maketitle

The main goal of this paper is to point out the close connection
between the description of an open quantum system (OQS) in the
Markovian approximation by the Lindblad equation  and the
description of its classical analog (CA), if it exists, by
equations of motion presented in a special form. To settle this
connection  in the easiest way one can use the correspondence
principle (CP) between the classical and the quantum Poisson
brackets. The principle was postulated by Dirac in his famous book
\cite{1}. Having this connection in hands we propose a new method
of quantization of non-Hamiltonian systems without memory.  The
procedure of using of the method consists in the following. Let us
assume that the quantum system of interest has the CA with known
equations of motion. We also imply that these equations can be
presented in a special form prescribed by the method - the form
allowing the quantization (FAQ). In this case, on the base of the
CP one can formulate an efficient recipe for finding the concrete
form of the operators that give the quantum description of this
system in the framework of the Lindblad equation. Below, we  call
such a procedure the semiclassical quantization of the dynamical
systems (without memory). We should emphasize that, to a certain
extent, the method of quantization proposed is a heuristic one.
Therefore, with reference to concrete examples, we demonstrate in
detail the specifics of its application. As such examples we
consider three popular dynamical systems:  1) the harmonic
oscillator with a friction; 2) the oscillator with a nonlinear
damping that simulates the mechanism of the emergence of the limit
cycle near the bifurcation point; and 3) the system of two
periodic rotators with a weak interaction that synchronizes their
oscillations. In our opinion, the latter example is in a direct
relation with the description of the quantum noise in a Josephson
junction with a strong damping. Besides, we give a simple
generalization of the method proposed that allows one to apply it
for the semiclassical quantization of open magnetic systems
described by the equation of the Landau-Lifshitz type with a
dissipation and under a pumping. Now, let us go to the details.

It is well-known that quantum mechanics in its original form (the
Schrodinger equation for the wave function $\Psi$ of the system:
$i\hbar \partial\Psi/\partial t =\hat{H}\Psi$) provides the
consistent description of the evolution of the state of a closed
microscopic system. Such a description is extended automatically
to the systems with a given dependence of their energy on time:
$\hat{H}=\hat{H}(t)$. If a closed quantum system has the classical
analog, the substitution of the semiclassical wave function
$\Psi=A \exp(i S_{cl}/\hbar)$ into the Schr\"{o}dinger equation
allows one to obtain  the well-known Hamilton-Jacobi equation for
the action function $S_{cl}(q,t)$ (in the limit $\hbar\to 0$),
see, e.g. \cite{2}. In that way one turns from the quantum to the
classical description of a closed system. If the microscopic
system $S$ we are interested in is an open one, i.e. it interacts
with an environment (a reservoir $R$) that, as a rule, contains a
large number of degrees of freedom, the quantum mechanics yields
the following recipe for the description of the evolution of the
state of $S$. One should consider the composite system $SR$ that
includes $S$ and the reservoir $R$ (that contains all exterior
degrees of freedom with which the system $S$ interacts). The
evolution of the density matrix of the closed system
$\hat{\rho}_{SR}(t)$ is described by the quantum  Liouville - von
Neumann equation:
\begin{equation}\label{1}
  \frac{d \hat{\rho}_{SR}}{d t} =-\frac{i}{\hbar} [\hat{H}_{SR},
  \hat{\rho}_{SR}],
\end{equation}
where $\hat{H}_{SR}=\hat{H}_S+\hat{H}_R+\hat{H}_{int}$ is the
Hamiltonian of the closed system $SR$. In case when the solution
of Eq. (\ref{1}) is known, the desired density matrix
$\hat{\rho}_S(t)$ is found by the reduction of
$\hat{\rho}_{SR}(t)$ over the variables that describe the
reservoir: $\hat{\rho}_{S}(t)={\rm tr}_R \hat{\rho}_{SR}(t)$. As a
rule, for real systems with interaction  this general recipe is
unrealizable. Therefore, most often, the procedure of reduction is
applied directly to the the equation (\ref{1}) and then one tries
to obtain a closed equation for $\hat{\rho}_{S}(t)$ using
approximations based on the presence of a small parameter in a
concrete physical system. On this way a large number of
interesting and important results for  quantum systems with
dissipation were obtained (see \cite{3} and references therein).
Nevertheless, such an approach is not a universal one, and in many
cases it cannot be realized. Regarding the standard approach one
should note that it looks rather "uneconomical" from the point of
view of the experiment, since the description of an open system
$S$ requires the full information about the system $SR$. In
connection with that,  an alternative approach based on the
quantum theory of open systems \cite{4} becomes more and more
popular during last years. In that approach,  one attempts to
describe the behavior of the OQS using only the information about
the system $S$. The approach is based on the general restrictions
imposed by the quantum mechanics on the evolution of
$\hat{\rho}_S(t)$: the linearity, the positivity, and the
conservation of the trace of $\hat{\rho}_S(t)$. One of the most
important results obtained in the framework of this approach is
the Lindblad equation. This equation realizes the quantum scheme
of the description of the evolution of the OQS in the Markovian
approximation. The Lindblad equation for the evolution of the
density matrix of OQS $\hat{\rho}_S(t)$ has the following form
\cite{5}:
\begin{equation}\label{2}
  \frac{d \hat{\rho}_S}{ d t} =-\frac{i}{\hbar} [\hat{H},
  \hat{\rho}_S]+\sum_{j=1}^N \left\{[\hat{R}_j\hat{\rho}_S,
  \hat{R}_j^+]+[\hat{R}_j, \hat{\rho}_S
  \hat{R}_j^+]\right\}.
\end{equation}

The first term in the right hand side of Eq. (\ref{2}) describes
the internal (Hamilton) dynamics of $S$, while the other terms
model its interaction with the reservoir $R$ ($N$ is the number of
channels of the interaction of $S$ with $R$). We emphasize that
the information determined in the standard approach by the
structure of the reservoir and by the type of the interaction
between $S$ and $R$ is included now in the operators $\hat{R}_j$,
$\hat{R}_j^+$ that acts in the space of states of the system $S$.

As an example, let us consider the situation when $N=1$ and the
operator $\hat{R}$ is the Hermitian one
$\hat{R}=\hat{R}^+=\hat{O}$. It turns out that in this important
special case Eq. (\ref{2}) describes the decoherence of the
quantum system interacting with the "meter" (macroscopic or
mesoscopic one) that measures continuously the observable
$\hat{O}$, see \cite{6}.

Let us assume now that there exists a quantum system $S$ (of the
Markov type) which is described by Eq. (\ref{2}) with some
$\hat{H}$, $\hat{R}_j$ and $\hat{R}_j^+$, and the system has the
classical analog. The question emerges: how can one find the
dynamics, i.e. the equations of motion,   of that CA from the
Lindblad equation for $S$? The answer on this question can be
obtained with the help of the correspondence principle formulated
by Dirac in \cite{1}. According to Dirac, the quantum Poisson
bracket (commutator) of two operators $\hat{A}$ and $\hat{B}$ is
connected with the Poisson bracket $\{A,B\}$ of their CA's -
$A(q_r,p_r)$ and $B(q_r,p_r)$ by the relation
\begin{equation}\label{3}
  [\hat{A},\hat{B}]=i\hbar \{A,B\} ,
\end{equation}
where, by definition,
\begin{equation}\label{4}
  \{A,B\} =\sum_r \frac{\partial A}{\partial q_r} \frac{\partial
  B}{\partial p_r} - \frac{\partial A}{\partial p_r} \frac{\partial
  B}{\partial q_r}.
\end{equation}
For the simplest operators $\hat{A}$ and $\hat{B}$ (for instance,
$\hat{A}=\hat{q}_r$ and $\hat{B}=\hat{p}_r$) the relation
(\ref{3}) is exact one. In other cases it is fulfilled
approximately, at least, in the first order in $\hbar$. Let us use
this CP for finding the classical analog of the Lindblad equation.
We imply that in the limit $\hbar \to 0$  the distribution
function $f_S(q_r,p_r)$ in the phase space of the system $S$
corresponds to the density matrix $\hat{\rho}_S$. We will consider
the simplest case, when $S$ has one degree of freedom and $N$, the
number of channels of the interaction of $S$ and $R$, is equal to
one. Let the operators  $\hat{H}$, $\hat{R}$ and $\hat{R}^+$ in
Eq. (\ref{2}) be the functions of the operators $\hat{q}$ and
$\hat{p}\equiv (\hbar/i)
\partial/\partial q$. Then, according to Eqs. (\ref{2}) and
(\ref{3}), one can write  the following  equation for the
distribution function of CA of the system S:
\begin{equation}\label{5}
  \frac{\partial f_S}{\partial t}=\{ H, f_S \}+i \hbar \{R f_S,
  \bar{R}\}+ i \hbar\{R, f_S \bar{R}\},
\end{equation}
where $H(p,q)$, $R(p,q)$ and $\bar{R}(p,q)$ are the classical
analogs of $\hat{H}$, $\hat{R}$ and $\hat{R}^+$, correspondingly.
Using the definition (\ref{4}) and differentiating explicitly, we
reduce Eq. (\ref{5}) to the following form:
\begin{equation}\label{6}
\frac{\partial f_S}{\partial t}+ A(q,p) \frac{\partial
f_S}{\partial q} + B(q,p) \frac{\partial f_S}{\partial p} +C(q,p)
f_S=0,
\end{equation}
where the notations
\begin{eqnarray}\label{7}
  A(q,p)= \frac{\partial H}{\partial p}+i\hbar \left(\bar{R}
  \frac{\partial R}{\partial p} -R \frac{\partial \bar{R}}{\partial
  p}\right), \cr
B(q,p)= -\frac{\partial H}{\partial q}+i\hbar \left(R
  \frac{\partial \bar{R}}{\partial q} -\bar{R} \frac{\partial R}{\partial
  q}\right),\cr
  C(q,p) =2 i \hbar \left( \frac{\partial R}{\partial
  p}\frac{\partial\bar{R}}{ \partial q} -  \frac{\partial R}{\partial
  q}\frac{\partial\bar{R}}{ \partial p}\right)
\end{eqnarray}
are used.

The first order  partial differential equation (\ref{6}) describes
the motion of the ensemble of systems (the phase liquid) in the
space $\Gamma_S\equiv (q,p)$ of the system $S$. It is known (see
\cite{7}) that the phase points (elements of the ensemble) move
along the characteristics of Eq. (\ref{6}) according to the
equations of motion:
\begin{eqnarray}\label{8}
  \dot{q} \equiv A(q,p)=\frac{\partial H}{\partial p}+i\hbar \left(\bar{R}
  \frac{\partial R}{\partial p} -R \frac{\partial \bar{R}}{\partial
  p}\right), \cr
  \dot{p}\equiv B(q,p)= -\frac{\partial H}{\partial q}+i\hbar \left(R
  \frac{\partial \bar{R}}{\partial q} -\bar{R} \frac{\partial R}{\partial
  q}\right).
\end{eqnarray}

Now, let us compute the divergence of the phase velocity ${\bf
v}=(\dot{q},\dot{p})$. By definition
\begin{equation}\label{9}
  {\rm div} {\bf v}= \frac{\partial \dot{q}}{\partial q}
  +\frac{\partial \dot{p}}{\partial p}=2 i \hbar \left( \frac{\partial \bar{R}}{\partial
  q}\frac{\partial R}{ \partial p} -  \frac{\partial R}{\partial
  q}\frac{\partial\bar{R}}{ \partial p}\right)
\end{equation}
Using the relations (\ref{8}) and (\ref{9}) we find that Eq.
(\ref{6}) is reduced to the standard form of the continuity
equation of motion for the phase liquid in the space $\Gamma_S$
\begin{equation}\label{10}
  \frac{\partial f_S}{\partial t} + {\rm div} (f_S {\bf v})=0.
\end{equation}

Thus, with the help of the correspondence principle (\ref{3}) we
arrive to the conclusion that in the limit $\hbar \to 0$  the
Lindblad equation (\ref{2}) turns to the continuity equation for
the distribution function $f_S$ for the CA of the system $S$. The
equations of motion for that CA  (Eq. (\ref{8})) follow directly
from Eq. (\ref{2}). This is the main result of the paper and we
will use it for solving the inverse problem: the obtaining of
quantum description of the dynamical system (without memory) from
the equations of motion for its CA.

It is convenient to present the equations of motion (\ref{8}) in
the form of one complex equation for the complex coordinate. Let
us define the complex variables $z$ and $z^*$:
\begin{equation}\label{11}
  z=\frac{\frac{q}{l_0}+\frac{i p l_0}{\hbar}}{\sqrt{2}}, \quad
  z^*=\frac{\frac{q}{l_0}-\frac{i p l_0}{\hbar}}{\sqrt{2}},
\end{equation}
where  $l_0$ is an arbitrary parameter with the dimension of the
length. It follows from the definition (\ref{11}) that for any
choice of $l_0$ the Poisson bracket of the dimensionless variables
$z$ and $z^*$ is
\begin{equation}\label{12}
  \{ z, z^*\}=-\frac{i}{\hbar} \{ q, p \}=-\frac{i}{\hbar}.
  \end{equation}
Using the system of equations of motion (\ref{8}) for $q$ and $p$,
we find the desired dynamical equation for $z$:
\begin{equation}\label{13}
  \frac{d z}{ d t} =- i \frac{\partial H} {\partial z^*}
  +\left(\bar{R} \frac{\partial R}{\partial z^*} - R
  \frac{\partial\bar{R}}{\partial z^*}\right).
\end{equation}
It is implied here that the functions $H(q,p)$, $R(q, p)$ and
$\bar{R}(q,p)$ in Eq. (\ref{13}) are expressed as the functions on
$z$ and $z^*$ with the help of relations inverse to Eq.
(\ref{11}):
\begin{equation}\label{14}
  q=\frac{l_0}{\sqrt{2}}(z^*+z), \quad p=\frac{i \hbar}{\sqrt{2}l_0}
  (z^*-z).
\end{equation}
The equation of motion for the complex coordinate $z$ written in
the form (\ref{13}) is referred as the one presented in the form
allowed the quantization (FAQ). Now, let us give the recipe (based
on the CP) for the transition from the description of the CA of
the system $S$ by the equation of motion to the quantum
description of $S$ by the Lindblad equation. The recipe proposed
consists of three consecutive steps.

1. The input dynamical equations should be presented in the FAQ
written for the complex coordinate $z$ in the form (\ref{13}).
Such a representation determines automatically the functions $H(z,
z^*)$, $R(z,z^*)$ and $\bar{R}(z,z^*)$ entered in the FAQ.

2. One finds the quantum analogs $\hat{H}$, $\hat{R}$ and
$\hat{R}^+$ of that functions. For that one should replace the
coordinates $z$ and $z^*$ to the corresponding operators $\hat{z}$
and $\hat{z}^+$. According to the CP (see (\ref{3})),
$[\hat{z},\hat{z}^+]=i\hbar\{z, z^*\}=1$. The latter relation
allows one to identify the operators $\hat{z}$ and $\hat{z}^+$
with the standard Bose operators $\hat{a}$ and $\hat{a}^+$, that
satisfy $[\hat{a},\hat{a}^+]=1$. We note that in the case when the
polynomial functions $H$, $R$ and $\bar{R}$ of $z$ and $z^*$
contain the terms of the form $z^k (z^*)^l$ the ordering of the
operators $\hat{a}$ and $\hat{a}^+$ should be fixed and necessary
symmetrization should be fulfilled under the transition to the
quantum analogs $\hat{H}$, $\hat{R}$ and $\hat{R}^+$. Below  we
demonstrate how this problem is resolved with reference to
concrete examples. Note, that the same problem emerges under the
quantization of the Hamilton systems, as well.

3. The operators  $\hat{H}$, $\hat{R}$ and $\hat{R}^+$ should be
substituted into the Lindblad equation (\ref{2}) for the evolution
of the density matrix. The correspondence principle guarantees
that the equation obtained by this method will give a correct
quantum description of the evolution of $S$, at least, with the
accuracy up to the first order in $\hbar$. Therefore, the
procedure described can be referred as the method of the
semiclassical quantization of dynamical systems (without memory).
We note that the  most delicate point of the method is the
presentation of the equations of motion in the FAQ - it is
difficult to formalize this point.

The rest of the paper is devoted to the study of concrete physical
systems that can be quantized by the method proposed.

\section*{Example A. Harmonic oscillator with  friction}

This simplest example of a quantum dissipative system was
considered more than once in the literature. In particular, in the
recent review by Mensky \cite{8} the equation of the Lindblad type
for the density matrix of such an oscillator was obtained by the
method of restricted Feynman integrals (the method of quantum
corridors). Let us show how simple the derivation of this equation
(up to the notations) by the method proposed here. The equation of
motion for the harmonic oscillator with friction has the
well-known form:
\begin{equation}\label{A1}
  m \ddot{q}+2\gamma \dot{q} + m\omega_0^2 q = 0,
\end{equation}
where $\gamma$ is the friction coefficient, $\omega_0$, the
oscillatory frequency, and m, its mass. Eq. (\ref{A1}) can be
presented as the system of two first order equations:
\begin{eqnarray}\label{A2}
  \dot{q}=\frac{p}{m}, \cr
  \dot{p}=-\frac{2\gamma}{m} p - m \omega_0^2 q .
\end{eqnarray}
Let us consider the complex coordinate $z=(q/l_0+i p
l_0/\hbar)/\sqrt{2}$. In this problem it is convenient to choose
the oscillatory length as the parameter
$l_0=\sqrt{\hbar/m\omega_0}$. At such $l_0$, using Eq. (\ref{A2}),
we obtain the equation of motion for $z$ in the form:
\begin{equation}\label{A3}
  \frac{d z} {d t} = -i \omega_0 z - \lambda (z-z^*),
\end{equation}
where the notation $\lambda\equiv\gamma/m$ is introduced.

Let us show that Eq. (\ref{A3}) can be presented in the FAQ. We
define two functions of the variables $z$ and $z^*$: the real one
$H=\hbar \omega_0 |z|^2 +i \hbar \lambda [(z*)^2-z^2]/2$ and the
complex one $R =\sqrt{\lambda}(z\cosh u  - z^*\sinh u)$, where $u$
is an arbitrary parameter. Then we check directly that the right
hand side of Eq. (\ref{A4}) can be written in the form
\begin{equation}\label{A4}
  - i \omega_0 z - \lambda  (z-z^*)= - \frac{i}{\hbar}
  \frac{\partial H}{\partial z^*} + \left(\bar{R} \frac{\partial
  R}{\partial z^*}- R \frac{\partial\bar{R}}{\partial z^*}\right).
\end{equation}
Therefore, in accordance with the method proposed, the quantum
description of the harmonic oscillator with friction is given by
the Lindblad equation
\begin{equation}\label{A6}
  \frac{d \hat{\rho}}{ d t} =-\frac{i}{\hbar} [\hat{H},
  \hat{\rho}]+ [\hat{R}\hat{\rho},
  \hat{R}^+]+[\hat{R}, \hat{\rho}\hat{R}^+],
\end{equation}
where the quantum analogs of the functions $H(z, z^*)$ and
$R(z,z^*)$ are determined as
\begin{eqnarray}\label{A7}
  \hat{H}=\hbar \omega_0 \hat{a}^+ \hat{a} + \frac{i \hbar \lambda}{2}[(\hat{a}^+)^2
  -\hat{a}^2],  \cr
  \hat{R} =\sqrt{\lambda} (\hat{a}\cosh u -\hat{a}^+\sinh u ).
\end{eqnarray}
Equation (\ref{A6}) together with the expressions for $\hat{H}$
and $\hat{R}$ (Eq. (\ref{A7})) coincides with the result Eq. (51)
of Ref. \cite{8} up to the notations. Using Eqs. (\ref{A6}) and
(\ref{A7}), one can derive the equations of motion for the first
($\overline{\hat{q}}$, $\overline{\hat{p}}$) and the second
($\overline{\hat{q}^2}$, $\overline{\hat{p}^2}$,
$\overline{\hat{q}\hat{p}+\hat{p}\hat{q}}$)  momenta of  $\hat{q}$
and $\hat{p}$ - the basic physical quantities of the system $S$.
But we will not do that here, since Ref. \cite{8} contains such a
derivation and the subsequent analysis (including a discussion on
the physical meaning of the parameter $u$). We have considered the
harmonic oscillator with only one aim: to illustrate all the
specifics of the method of quantization on the example of the
simplest model. Let us now turn to more interesting examples. To
simplify the notations, we will use the system of units with
$\hbar=1$. Also, the dimensionless coordinates $x\equiv q/l_0$ and
$y\equiv p l_0/\hbar$ will be used instead of the variables $q$
and $p$. Consequently, the complex coordinate $z$ in the FAQ is
defined as $z=(x+i y)/\sqrt{2}$.

\section*{Example B. Quantum analog of the bifurcation of the emergence of the cycle}

Let us show how the quantum description of the oscillator with a
nonlinear damping can be given by the method proposed. The CA of
such an oscillator simulates the mechanism of a smooth emergence
of auto-oscillations near the bifurcation point. Similar
bifurcation of the emergence of the cycle is typical  for the
situation when an equilibrium point losses its stability under a
variation of the parameters of the dynamical system. In such a
situation the growing up  of small oscillations may switch the
system to a new stationary state that corresponds to the closed
trajectory (limit cycle). The following system of equations gives
the mathematical model of the behavior of the dynamical system
near the bifurcation of the emergence of the cycle (see \cite{9}):
\begin{eqnarray}\label{B1}
  \frac{d x} {d t} =\omega y +\lambda x -\mu x (x^2+y^2), \cr
  \frac{d y} {d t}= -\omega x +\lambda y -\mu y (x^2+y^2).
\end{eqnarray}
The system of the equations of motion Eq. (\ref{B1}) can be
written in the complex form:
\begin{equation}\label{B2}
  \frac{d z}{d t}= - i\omega z +\lambda z - 2\mu z |z|^2.
  \end{equation}
Equation (\ref{B2}) can be presented in the FAQ. To show this we
introduce the functions $H=\hbar \omega z^* z$,
$R_1=\sqrt{\lambda} z^*$ and $R_2=\sqrt{\mu} z^2$. Then we check
directly that the right hand side of Eq. (\ref{B2}) can be
rewritten in the form
\begin{eqnarray}\label{B3}
  -i \omega z  +\lambda z - 2 \mu z |z|^2= -i \frac{\partial
  H}{\partial z^*} \cr+\left(\bar{R}_1 \frac{\partial
  R_1}{\partial z^*}- R_1 \frac{\partial\bar{R}_1}{\partial
  z^*}\right)
  +\left(\bar{R}_2 \frac{\partial
  R_2}{\partial z^*}- R_2 \frac{\partial\bar{R}_2}{\partial
  z^*}\right).
\end{eqnarray}
According to the recipe of quantization described above, the
Lindblad equation for the evolution of the density matrix of the
quantum analog of the system (\ref{B1}) can be written in the form
\begin{eqnarray}\label{B4}
 \frac{d \hat{\rho}}{ d t} =-i [\hat{H},
  \hat{\rho}]+ [\hat{R}_1\hat{\rho},
  \hat{R}_1^+]+[\hat{R}_1, \hat{\rho}\hat{R}_1^+]\cr +
  [\hat{R}_2\hat{\rho},
  \hat{R}^+_2]+[\hat{R}_2, \hat{\rho}\hat{R}^+_2]
\end{eqnarray}
with $\hat{H}=\omega \hat{a}^+ \hat{a}$, $\hat{R}_1=\sqrt{\lambda}
\hat{a}^+$, and $\hat{R}_2=\sqrt{\mu} \hat{a}^2$.

From physical reasons it is clear that the motion of the system
(\ref{B1}) along the limit cycle should correspond to the
stationary solution of Eq. (\ref{B4}) $\hat{\rho}_{st}$, such that
$d \hat{\rho}_{st}/ d t =0$. By analogy with the classical
solution, for which $|z|^2\equiv const$, one can expect that
$\hat{\rho}_{st}$ will commute with the particle number operator
$\hat{N}=\hat{a}^+ \hat{a}$. Therefore, we will seek the
stationary solution of Eq. (\ref{B4}) in the form
$\hat{\rho}_{st}=\sum_{n=0}^\infty |n\rangle \rho_n \langle n |$,
where $|n\rangle$ are the eigenfunctions of the operator
$\hat{N}$. Positive coefficients $\rho_n$ satisfy the additional
normalization condition $\sum_{n=0}^\infty \rho_n=1$. Substituting
this expansion into Eq. (\ref{B4}) we obtain the system of
recurrent relations for the coefficients $\rho_n$:
\begin{eqnarray}\label{B5}
    2\lambda [ n \rho_{n-1} - (n+1) \rho_n]+\cr 2 \mu
    [(n+2)(n+1)\rho_{n+2}- n (n-1)\rho_n]=0.
\end{eqnarray}
It is convenient to introduce the generating function for the
coefficients $\rho_n$. By definition $G(u)\equiv \sum_{n=0}^\infty
\rho_n u^n$. One can check by the direct substitution that the
system (\ref{B5}) is equivalent to the following second order
differential equation for the function $G(u)$:
\begin{equation}\label{B6}
    (1+u) \frac{ d ^2 G}{ d u^2}-\nu u \frac{ d G } {d u} -\nu
    G(u)=0.
\end{equation}
In Eq. (\ref{B6}) the notation $\nu=\lambda/\mu$ is used.

Let us transform Eq. (\ref{B6}) to the canonical form. To do this
we introduce the auxiliary variable $v=\nu(1+u)$. Then $d/ d u =
\nu d/d v$, and  Eq. (\ref{B6}) becomes
\begin{equation}\label{B7}
v \frac{ d ^2 G}{ d v^2}+(\nu-v) \frac{ d G } {d v} -
    G(v)=0.
\end{equation}
Eq. (\ref{B7}) coincides with the equation for the confluent
hypergeometric function $y=\Phi(a,c,x)$:
\begin{equation}\label{B8}
x \frac{ d ^2 y}{ d x^2}+(c-x) \frac{ d y } {d x} -
    a y(x)=0
\end{equation}
(see \cite{10}) if one sets $c=\nu$ and $a=1$.

Therefore, the solution of Eq. (\ref{B6}) that satisfies all
conditions of the problem can be written in the form
\begin{equation}\label{B9}
    G(u)=\frac{\Phi(1,\nu,\nu(1+u))}{\Phi(1,\nu, 2 \nu)}.
\end{equation}
In the solution Eq. (\ref{B9}) we take into account that the
normalization condition $\sum_{n=0}^\infty \rho_n=1$ is equivalent
to the relation $G(1)=1$ for the generating function $G(u)$.

Differentiating the expression Eq. (\ref{B9}) one can find the
average values of all physical quantities of interest in the
stationary state. For instance, the average number of quanta
$\bar{n}$ generated in the stationary state is given by the
relation
\begin{equation}\label{B10}
    \bar{n}=\sum_{n=0}^\infty n \rho_n =\left. \frac{d G(u)}{d u}
    \right|_{u=1},
\end{equation}
and the quantity $\overline{n^2}-\overline{n}=\overline{(n-1)n}$
is connected with the second derivative of $G(u)$ by the relation
\begin{equation}\label{B11}
\overline{n^2}-\overline{n}=\left. \frac{d^2 G(u)}{d u^2}
    \right|_{u=1}.
\end{equation}

To characterize the properties of the distribution of the number
of quanta $\{\rho_n\}$ we will use the Mandel parameter $Q$
defined as (see \cite{11})
\begin{equation}\label{B12}
Q=\frac{\overline{n^2}-\overline{n}^2}{\overline{n}}-1=\frac{\overline{n(n-1)}}{\overline{n}}
-\overline{n}.
\end{equation}
The distributions for which $Q>0$ are called the super-Poisson
ones, while the distribution with $Q<0$, the sub-Poisson ones. The
Poisson distribution corresponds to $Q=0$. Using the properties of
the confluent hypergeometric function, one can easily find that at
$\nu=1$ the  expression Eq. (\ref{B9}) corresponds to the Poisson
distribution with $\overline{n}=1$, at $\nu>1$ it describes the
super-Poisson distribution, and at $\nu<1$, the sub-Poisson
distribution. We note also that for integer $\nu$ the generating
function $G(u)$ can be presented through elementary functions. For
instance, at $\nu=1$ it reads as $G(u)=e^{u-1}$, at  $\nu=2$,
$G(u)=(2/\sinh 2) (\sinh(1+u)/(1+u)) e^{u-1}$ etc. Thus, the
method proposed allows one to describe in  detail the stationary
state of OQS which classical analog simulates the bifurcation that
occurs very frequently in the nonlinear mechanics.

\section*{Example C. Quantum analog of  the phase synchronization of two
auto-oscillating rotators}

In this part of the paper we consider the specifics of application
of the method for the problem of quantization of dynamical systems
with several degrees of freedom. While such a generalization looks
quite obvious, nevertheless, additional interesting opportunities
emerge, that will be analyzed  in detail. As a distinctive example
that is of interest by itself, we consider  the dynamical system
of two interacting auto-oscillating rotators. In the case when
their frequencies (in the absence of the interaction) are rather
close to each other the switching on the interaction may result in
the effect of synchronization, i.e. the locking of the frequencies
and the phases in the bound system. The classical description of
the effect of synchronization is well-known \cite{12}. Here we are
interested in the possibility of its quantum description. We
restrict the consideration by the case of weak interaction between
the periodic oscillators, when the method of phase dynamics (see
\cite{12}) can be used for their description.  For the simplest
form of the coupling between the oscillators this method yields
the following  system of equations of motion for their phases
$\varphi_1$ and $\varphi_2$
\begin{eqnarray}\label{C1}
    \frac{ d \varphi_1} {d t} =\omega_1+ a \sin
    (\varphi_2-\varphi_1),\cr
 \frac{ d \varphi_2} {d t} =\omega_2+ a \sin
    (\varphi_1-\varphi_2),
\end{eqnarray}
where $\omega_1$ $(\omega_2)$ is the frequency of motion the first
(second) oscillator in its limit cycle,  and $a$ is the coupling
constant that determines the strength of the interaction between
the oscillators.

It is implied in this approximation that each oscillator (rotator)
moves along the limit cycle with a constant amplitude, i.e., the
additional restrictions for the system Eq. (\ref{C1}) ($d r_1/d t=
d r_2/ d t=0$) are assumed. Taking into account these restrictions
explicitly, one can write the system Eq. (\ref{C1}) in the
following equivalent form:
\begin{eqnarray}\label{C2}
    \frac{ d \varphi_1} {d t} =\omega_1+ \lambda r_1 r_2 \sin
    (\varphi_2-\varphi_1),\cr
 \frac{ d \varphi_2} {d t} =\omega_2+ \lambda r_1 r_2 \sin
    (\varphi_1-\varphi_2),\cr
    \frac{ d r_1}{ d t}=\frac{d r_2} {d t}=0.
\end{eqnarray}
The system Eq. (\ref{C2}) can be presented in the form of two
dynamical equations for the complex coordinates $z_1=(x_1+ i
y_1)/\sqrt{2}=r_1 e^{i\varphi_1}/\sqrt{2}$ and $z_2=(x_2+ i
y_2)/\sqrt{2}=r_2 e^{i\varphi_2}/\sqrt{2}$:
\begin{eqnarray}\label{C3}
    \frac{ d z_1} {d t} = i \omega_1 z_1 + \lambda z_1 (z_1^* z_2-z_2^* z_1),\cr
 \frac{ d z_2} {d t} =i \omega_2 z_2+ \lambda z_2(z_2^* z_1-z_1^*
 z_2).
\end{eqnarray}
Indeed, if one substitutes the expressions for $z_1$ and $z_2$
into Eq. (\ref{C3}) and then separates the real and the imaginary
parts, one obtains just the system (\ref{C3}). Let us show that
Eq. (\ref{C3}) can be presented in the FAQ. To do  this we choose
the following functions $H$ and $R$:
\begin{eqnarray}\label{C4}
    H=-\omega_1 |z_1|^2- \omega_2|z_2|^2+\frac{i\lambda
    |z_1|^2}{2}(z_1^* z_2-z_2^* z_1)\cr+\frac{i\lambda
    |z_2|^2}{2}(z_2^* z_1-z_1^* z_2),\cr
    R=\frac{\sqrt{\lambda}}{2}\left(
    |z_1|^2-|z_2|^2+z_2^*z_1-z_2z_1^*\right).
\end{eqnarray}
One can check directly that the right hand sides of the equations
of the system (\ref{C3}) are expressed as
\begin{eqnarray}\label{C5}
i \omega_1 z_1 + \lambda z_1 (z_1^* z_2-z_2^* z_1)=-i
\frac{\partial H}{\partial z_1^*}\cr+\left(\bar{R} \frac{\partial
R}{\partial z_1^*}-R \frac{\partial \bar{R}}{\partial
z_1^*}\right),\cr i \omega_2 z_2+ \lambda z_2(z_2^* z_1-z_1^*
z_2)= -i \frac{\partial H}{\partial z_2^*}\cr+\left(\bar{R}
\frac{\partial R}{\partial z_2^*}-R \frac{\partial
\bar{R}}{\partial z_2^*}\right).
\end{eqnarray}
Then, following the method of quantization proposed, one can pass
to the  corresponding Lindblad equation in which the operators
$\hat{H}$ and $\hat{R}$ are obtained from the functions (\ref{C4})
by the substitution $z_1\to \hat{a}_1$,  $z_1^*\to \hat{a}_1^+$,
$z_2\to \hat{a}_2$, $z_2^*\to \hat{a}_2^+$. But,  there is another
formulation of this problem, that is  especially useful for the
quantization of open magnetic systems. Let us describe in short
this formulation.

Let the dynamical systems with two degrees of freedom be
determined by its equations of motion presented in FAQ with some
functions $H$ and $R$
\begin{eqnarray}\label{C6}
\frac{ d z_1}{ d t}=-i \frac{\partial H}{\partial
z_1^*}+\left(\bar{R} \frac{\partial R}{\partial z_1^*}-R
\frac{\partial \bar{R}}{\partial z_1^*}\right),\cr \frac{ d z_2}{
d t}= -i \frac{\partial H}{\partial z_2^*}+\left(\bar{R}
\frac{\partial R}{\partial z_2^*}-R \frac{\partial
\bar{R}}{\partial z_2^*}\right).
\end{eqnarray}
Instead of the complex coordinates $z_1$ and $z_2$ we introduce
three real valued variables $l_x=(z_1^*z_2+z_2^*z_1)/2$,
$l_y=i(z_2^* z_1-z_1^* z_2)/2$, and $l_z=(|z_1|^2-|z_2|^2)/2$. We
imply that the Poisson brackets for the coordinates $z_1$,
$z_1^*$, $z_2$, and $z_2^*$ satisfy the relations $\{z_\alpha,
z_\beta\}=0$, $\{z_\alpha^*, z_\beta^*\}=0$, and $\{z_\alpha,
z_\beta^*\}=-i\delta_{\alpha\beta}$, where $\alpha,\beta=1,2$.
Then it follows from the definition of $l_x$, $l_y$ and $l_z$ that
the Poisson brackets for the components of the vector ${\bf
l}=(l_x,l_y,l_z)$ have the form $\{l_i,l_k\}=\epsilon_{ijk}l_k$
($i,j,k=1,2,3$). It allows one to identify the vector ${\bf l}$
with the angular momentum of the system considered. If $H$ and $R$
in Eq. (\ref{C6}) can be presented as functions of $l_x$, $l_y$
and $l_z$, then, substituting the variables and doing a simple
algebra, we find that the system Eq.(\ref{C6}) is equivalent to
the following equation of motion for the vector ${\bf l}$:
\begin{equation}\label{C7}
    \frac{d {\bf l}}{d t} = - \left({\bf l} \times \frac{ \delta
    H}{\delta{\bf l}}\right)+ i R \left({\bf l} \times \frac{ \delta
    \bar{R}}{\delta{\bf l}}\right) + c.c.
\end{equation}
Eq. (\ref{C7}) can be considered as a generalization of the
standard equation of motion for the magnetic momentum
(Landau-Lifshitz equation) for the case of the momentum
interacting with the environment. One can see from Eq. (\ref{C7})
that such an equation is valid under the condition of conservation
of the total magnetic momentum under the motion. Physically, it
means that the interaction of the open system with the environment
is invariant with respect to rotations of the system. The
representation of the equations of motion for the magnetic
momentum in the form (\ref{C7}) with given $H({\bf l})$, $R({\bf
l})$ and $\bar{R}({\bf l})$ allows one to fulfill the obvious
transition to the quantum description on the base of the Lindblad
equation. To do this one should find the operators $\hat{H}$,
$\hat{R}$ and $\hat{R}^+$, the quantum analogs of the functions
$H$, $R$ and $\bar{R}$, correspondingly. As before, the procedure
of finding of that operators is determined by the correspondence
principle: one should replace the variables $l_x$, $l_y$ and $l_z$
in the expressions for $H$, $R$ and $\bar{R}$ to the operators
$\hat{l}_x$, $\hat{l}_y$ and $\hat{l}_z$ that satisfy the
commutation relations $[\hat{l}_i,
\hat{l}_j]=i\epsilon_{ijk}\hat{l}_k$. Then, if necessary, the
expressions obtained should be ordered and symmetrized.

Let us now show how this variant of the quantization method works
in the problem we are interested in -  the problem on the
synchronization of two interacting rotators. First of all, we note
that the quantities $H$, $R$ and $\bar{R}$ in Eq. (\ref{C4}) are
expressed as functions of $l_x$, $l_y$ and $l_z$ as follows
\begin{eqnarray}\label{C8}
    H=-(\omega_1-\omega_2) l_z- 2 \lambda l_y l_z,\cr
    R=\sqrt{\lambda}(l_z-i l_y).
\end{eqnarray}
We omit the term $-(\omega_1+\omega_2)(|z_1|^2+|z_2|^2)/2$ in the
expression for $H$ in Eq. (\ref{C8}), because this term is the
integral of motion of the equations (\ref{C3}).

The quantum description of the system we are interested in is
given by the following Lindblad equation:
\begin{equation}\label{C9}
  \frac{d \hat{\rho}}{ d t} =-\frac{i}{\hbar} [\hat{H},
  \hat{\rho}]+ [\hat{R}\hat{\rho},
  \hat{R}^+]+[\hat{R}, \hat{\rho}\hat{R}^+],
\end{equation}
in which $\hat{H}$ and $\hat{R}$ are the quantum analogs of the
functions $H$ and $R$ (Eq.(\ref{C8})):
\begin{eqnarray}\label{C10}
    H=-(\omega_1-\omega_2) \hat{l}_z-  \lambda \left(\hat{l}_y \hat{l}_z+
    \hat{l}_z \hat{l}_y\right),\cr
    \hat{R}=\sqrt{\lambda}\left(\hat{l}_z-i \hat{l}_y\right).
\end{eqnarray}
Using the Lindblad equation (\ref{C9}) one can find the equation
of motion for the average value of an arbitrary observable
$\hat{A}$: $\overline{\hat{A}}\equiv {\rm tr} \hat{\rho} \hat{A}$.
The direct calculation of $d \overline{\hat{A}}/d t$ yields the
following result:
\begin{equation}\label{C11}
    \frac{ d \overline{\hat{A}}}{ d t} = - i
    \overline{[\hat{A},\hat{H}]}+\overline{\hat{R}^+[\hat{A},\hat{R}]}+
    \overline{[\hat{R}^+, \hat{A}]\hat{R}}.
\end{equation}
Using Eq. (\ref{C11}) we obtain  the system of equations of motion
for the first order momenta $\overline{\hat{l}_x}$,
$\overline{\hat{l}_y}$ and $\overline{\hat{l}_z}$:
\begin{eqnarray}\label{C12}
&&    \frac{d \overline{\hat{l}_x}}{ d t} = - 2\lambda
    \overline{\hat{l}_x}+ 4 \lambda \overline{\hat{l}_y^2}+\delta
    \overline{\hat{l}_y},\cr
&&\frac{d \overline{\hat{l}_y}}{ d t}=
-\lambda\overline{\hat{l}_y}- 2 \lambda(
\overline{\hat{l}_x\hat{l}_y+\hat{l}_y\hat{l}_x})-\delta\overline{\hat{l}_x},\cr
&&\frac{d \overline{\hat{l}_z}}{ d t} =- \lambda
\overline{\hat{l}_z}.
\end{eqnarray}
In Eq. (\ref{C12}) we introduce the notation
$\delta\equiv\omega_1-\omega_2$ for the difference of the
frequencies of two rotators.

One can see from Eq. (\ref{C12}) that the equations of motion for
the first order momenta contain the second order momenta
$\overline{\hat{l}_y^2}$ and
$\overline{\hat{l}_x\hat{l}_y+\hat{l}_y\hat{l}_x}$. Writing the
equations of motion for the second order momenta
\begin{eqnarray}\label{C13}
    &&\frac{ d \overline{\hat{l}_y^2}} {d t} =  2 \lambda
    (\overline{\hat{l}_x^2}-\overline{\hat{l}_z^2})-\delta
    (\overline{\hat{l}_y\hat{l}_x+\hat{l}_x\hat{l}_y}),\cr
&&\frac{ d \overline{\hat{l}_z^2}} {d t} = -2 \lambda
\overline{\hat{l}_y(\hat{l}_x\hat{l}_y+\hat{l}_y\hat{l}_x)}\cr&&-
2 \lambda
\overline{(\hat{l}_x\hat{l}_y+\hat{l}_y\hat{l}_x)\hat{l}_y}+2
\lambda (\overline{\hat{l}_x^2}-\overline{\hat{l}_y^2}),\cr
&&\frac {d (\overline{\hat{l}_x\hat{l}_y+\hat{l}_y\hat{l}_x})}{ d
t} = 8 \lambda \overline{\hat{l}_y^3}- 3 \lambda
(\overline{\hat{l}_y\hat{l}_x^2+\hat{l}_x^2\hat{l}_y}) \cr && -
5\lambda (\overline{\hat{l}_y\hat{l}_x+\hat{l}_x\hat{l}_y})-
2\lambda \overline{\hat{l}_x\hat{l}_y\hat{l}_z} +2 \delta
(\overline{\hat{l}_y^2}-\overline{\hat{l}_x^2}) -\lambda
\overline{\hat{l}_y},
\end{eqnarray}
we note that they contain the third order momenta, etc. Thus,
further simplifications are needed for obtaining the solution of
the quantum problem. Below, we consider only stationary states of
the system (for which the left hand sides of Eqs. (\ref{C12}) and
(\ref{C13}) are zero) and specify the case of complete
synchronization $\delta=0$. To obtain the closed system of
equation for the first order momenta and the second order momenta
we decouple the third order momenta applying the approximation
frequently used in such problems (see \cite{13}):
$$
\overline{\hat{A}\hat{B}\hat{C}}\approx
\overline{\hat{A}\hat{B}}\cdot\overline{\hat{C}}+\overline{\hat{A}}
\cdot\overline{\hat{B}\hat{C}}+\overline{\hat{A}\hat{C}}\cdot\overline{\hat{B}}-
2
\overline{\hat{A}}\cdot\overline{\hat{B}}\cdot\overline{\hat{C}}.
$$
As a final result, we obtain the following system of equations for
the first and the second order  momenta
\begin{eqnarray}\label{C14}
   && \overline{\hat{l}_z}=0, \quad \overline{\hat{l}_x\hat{l}_y}=
    \overline{\hat{l}_y\hat{l}_x}=-\frac{\overline{\hat{l}_y}}{4},\cr
    &&\overline{\hat{l}_y^2}=\frac{\overline{\hat{l}_x}}{2}, \quad
    \overline{\hat{l}_x^2}=\overline{\hat{l}_z^2},\cr
    && 4 \left[ 2
    \overline{\hat{l}_y}\cdot\overline{\hat{l}_x\hat{l}_y}
    +\overline{\hat{l}_x}\cdot\overline{\hat{l}_y^2}
    -2\overline{\hat{l}_x}\cdot\left(\overline{\hat{l}_y}\right)^2\right]+
    (\overline{\hat{l}_y^2}-\overline{\hat{l}_x^2})=0,\cr
    && 8 \left[ 3\overline{\hat{l}_y^2}\cdot \overline{\hat{l}_y}-
    2\left(\overline{\hat{l}_y}\right)^3\right]-8\Big[2\overline{\hat{l}_x\hat{l}_y}
    \cdot\overline{\hat{l}_x}+\overline{\hat{l}_y}\cdot\overline{\hat{l}_x^2}\cr &&-
    2\overline{\hat{l}_y}\left(\overline{\hat{l}_x}\right)^2\Big]
    - 10 \overline{\hat{l}_x\hat{l}_y}-\overline{\hat{l}_y}=0.
\end{eqnarray}
It follows from the last equation of the system (\ref{C14}) that
$\overline{\hat{l}_y}=0$. This relation is valid also for the CA.
Let us remind that in the classical limit ($\hbar\to 0$)
$\hat{l}_y\to l_y=r_1 r_2 \sin(\varphi_2-\varphi_1)\to 0$, since
for the case of complete synchronization the phase difference
$\varphi=\varphi_2-\varphi_1\to 0$. But, in difference with the
CA, where $\overline{l_y^2}=r_1^2r_2^2\overline{\sin^2\varphi}$ is
also approaches zero, in the quantum case
$\overline{\hat{l}_y^2}\neq 0$ for the stationary state, and this
quantity can be the measure of the quantum noise of the system.
Let us find the value of $\overline{\hat{l}_y^2}$ using the system
(\ref{C14}) and the angular momentum conservation law
$\hat{l}_x^2+\hat{l}_y^2+\hat{l}_z^2\equiv
\hat{L}^2=l(l+1)=(N/2)(N/2+1)$, where $\hat{N}=
\hat{a}^+_1\hat{a}_1+\hat{a}_2^+ \hat{a}_2$ is the total number of
the excitations under the bosonic description of the system. Let
us notate $x\equiv \overline{\hat{l}_y^2}$. It follows from
(\ref{C14}) that $x=\overline{\hat{l}_x}/2$ and it satisfies the
simple equation
\begin{equation}\label{C15}
    8 x^2 +\frac{3 x}{2}-\frac{N^2}{8}-\frac{N}{4}=0.
\end{equation}
The solution of Eq. (\ref{C15}) is
\begin{equation}\label{C16}
   \overline{\hat{l}_y^2}=x=\frac{1}{8}\left[\left(N^2+2
    N+\frac{9}{16}\right)^{1/2}-\frac{3}{4}\right]\approx \frac{N}{8}.
\end{equation}
In  Eq. (\ref{C16}) we take into account that the semiclassical
approximation requires the fulfilment of the inequality $N\gg 1$.

Let us summarize the main conclusions that concern the latter
example.

1. The quantum description  of the non-Hamiltonian system
consisting of two rotators which oscillations be synchronized by
their mutual interaction is possible not only with the use of
standard Bose operators, but also in the representation, where the
components of the operator of the angular momentum $\hat {{\bf
l}}=(\hat{l}_x,\hat{l}_y,\hat{l}_z)$ be used as the generators.

2. Under the condition of complete synchronization
($\omega_1=\omega_2$) the level of the quantum noise in this
system is determined by the average value of the operator
$\hat{l}_y^2$.

3. In the stationary state $\overline{\hat{l}_y^2}\simeq N/8$,
where $N$ is the total number of the excitations under the bosonic
description of the system

It is necessary to note that the modification of the quantization
method used in the latter example is based on the Schwinger
representation \cite{14} for the angular momentum operator in
terms of two Bose operators.

To conclude this consideration, we discuss briefly the possibility
of the application of the results of the last section to the study
of quantum fluctuations in Josephson junctions with a strong
damping. Usually, the dynamics of such a junction is described by
the resistive model on the base of the following well-known
equation \cite{15}
\begin{equation}\label{C17}
    \frac{\hbar\dot{\varphi}}{2 e R}=I_{ex}-I_c \sin\varphi.
\end{equation}
In Eq. (\ref{C17}) the following notations are used:
$\varphi=\varphi_1-\varphi_2$ is the phase difference between the
superconducting banks of the junction, $I_{ex}$, the input current
applied to the banks, $R$, the Ohmic resistance of the junction,
$I_c$, its critical current. Comparing Eqs. (\ref{C17}) and
(\ref{C1}) we see that these equations coincide with each other up
to the notations. The regime of the complete synchronization
$\omega_1=\omega_2$ in the model of coupled rotators corresponds
to the case of zero input current $I_{ex}$ in the Josephson
junction model. As is easily seen, the parameter of the quantum
noise $\overline{\hat{l}_y^2}$ is proportional to the average
square of the bias voltage in the junction and this quantity can
be measured experimentally. It is also obvious that the number of
bosonic excitations
$\hat{N}=\hat{a}^+_1\hat{a}_1+\hat{a}^+_2\hat{a}_2$ in the
Josephson junction is equal to the number of Cooper pairs in its
banks. In the framework of the Ginzburg-Landau theory this
quantity is simply connected with the order parameter of the
superconductor \cite{15}.  The explicit form of the relation
(\ref{C16}) for the Josephson junction and a detailed analysis of
the quantum fluctuations by the method proposed will be given in
further publications.

I would acknowledge L.A.Pastur for the discussion of the results
of the paper and valuable comments.

%\section*{References}


\begin{thebibliography}{99}
\bibitem{1} P.A.M. Dirac, The Principles of Quantum Mechanics,
4-th ed., Clarendon Press, Oxford, 1958,
\bibitem{2} L.D.Landau and E.M.Lifshitz, Quantum Mechanics,
Pergamon Press, London, 1958.
\bibitem{3} U. Weiss, Quantum Dissipative Systems, World
Scientific, Singapore, 1999.
\bibitem{4} E.B. Davies, Quantum Theory of Open Systems, Academic
Press, London- New York, 1976.
\bibitem{5} G.Lindblad, Commun. Math. Physics {\bf 48}, 119
(1976).
\bibitem{6} M.B. Mensky, Phys-Usp. {\bf 41}, 923 (1998)[ Usp. Fiz.
Nauk {\bf 168}, 1019 (1998)].
\bibitem{7} R. Courant, Partial Differential Equations,
Interscience, New-York - London, 1962.
\bibitem{8} M.B. Mensky, Phys-Usp. {\bf 46}, 1163 (2003)[ Usp. Fiz.
Nauk {\bf 173}, 1199 (2003)].
\bibitem{9} V.I.Arnold, Geometrical Methods in the Theory of
Ordinary Differential Equations, 2-nd ed., Springer, New-York,
1988.
\bibitem{10} H. Bateman and A. Erdelyi (eds.), Higher
Transcendental Functions, v. 1, chap. 6, McGraw-Hill, New-York,
1953.
\bibitem{11} M.O. Scully and M.S. Zubairy, Quantum Optics,
Cambridge University Press, 1997.
\bibitem{12} A. Pikovsky, M. Rosenblum, J. Kurths,
Synchronization. An Universal Concept in Nonlinear Science,
Cambridge University Press, 2002.
\bibitem{13} A. Vardi, J. R. Anglin, Phys. Rev. Lett. {\bf 86},
568 (2001)
\bibitem{14} J. Schwinger, in: Quantum theory of angular momentum,
L.C. Biederharn and H. van Dam (eds.), Academic Press, New York,
1965, p. 229.
\bibitem{15} A. Barone, G. Paterno, Physics and Applications of
the Josephson effect, Wiley, New-York, 1982.
\end{thebibliography}
\end{document}